\documentclass{article}
\usepackage{geometry}
\usepackage{authblk}
\usepackage{amsmath}
\usepackage{amssymb}
\usepackage{graphicx}
\usepackage{dcolumn}
\usepackage{subfigure}
\usepackage{bm}
\usepackage{amsfonts}
\usepackage{color}
\usepackage{algorithm}
\usepackage{algorithmicx}
\usepackage{algpseudocode}
\usepackage{dsfont}

\begin{document}
\title{Quantum generative adversarial network for generating discrete distribution}

\author[1]{Haozhen Situ}
\author[2]{Zhimin He}
\author[3]{Yuyi Wang}
\author[,3]{Lvzhou Li\thanks{Corresponding author.\\ E-mail addresses: situhaozhen@gmail.com (H. Situ), zhmihe@gmail.com (Z. He), yuwang@ethz.ch (Y. Wang), lilvzh@mail.sysu.edu.cn (L. Li), zhengshg@pcl.ac.cn (S. Zheng).}}
\author[4,5]{Shenggen Zheng}

\affil[1]{\scriptsize College of Mathematics and Informatics, South China Agricultural University, Guangzhou 510642, China}
\affil[2]{School of Electronic and Information Engineering, Foshan University, Foshan 528000, China}
\affil[3]{School of Data and Computer Science, Sun Yat-Sen University, Guangzhou 510006, China}
\affil[4]{Center for Quantum Computing, Peng Cheng Laboratory, Shenzhen 518055, China}
\affil[5]{Institute for Quantum Science and Engineering, Southern University of Science and Technology, Shenzhen 518055, China}

\date{}
\maketitle

\begin{abstract}
Quantum machine learning has recently attracted much attention from the community of quantum computing. In this paper, we explore the ability of generative adversarial networks (GANs) based on quantum computing. More specifically, we propose a quantum GAN for generating classical discrete distribution, which has a classical-quantum hybrid architecture and is composed of a parameterized quantum circuit as the generator and a classical neural network as the discriminator. The parameterized quantum circuit only consists of simple one-qubit rotation gates and two-qubit controlled-phase gates that are available in current quantum devices. Our scheme has the following characteristics and potential advantages: (i) It is intrinsically capable of generating discrete data (e.g., text data), while classical GANs are clumsy for this task due to the vanishing gradient problem. (ii) Our scheme avoids the input/output bottlenecks embarrassing most of the existing quantum learning algorithms that either require to encode the classical input data into quantum states, or output a quantum state corresponding to the solution instead of giving the solution itself, which  inevitably  compromises the speedup of the quantum algorithm.  (iii) The  probability distribution  implicitly given by data samples  can be loaded into a quantum state, which may be useful for some further applications.
\end{abstract}

\section{Introduction}

Designing explicit algorithms for artificial intelligence problems, \emph{e.g.}, image and speech recognition, is very difficult or even impossible. Machine learning tries to solve these problems by parameterizing structured models and using empirical data to learn the parameters. There are two main classes of machine learning tasks, supervised learning and unsupervised learning. The goal of supervised learning is to build the relation between given data samples and their labels, while unsupervised learning aims at discovering the intrinsic patterns, properties or structures of unlabeled data samples. Compared with advanced supervised learning techniques, unsupervised learning is more intractable and challenging,
because it requires one to efficiently represent, learn and sample from high-dimensional probability distributions \cite{DLbook}.

Generative modeling is an important unsupervised learning subject. Generative models aim to emulate the distribution of training data and generate new instances accordingly, and a number of deep generative models have been proposed that are capable to create novel human-level art. Have you ever imagined that you could paint like da Vinci or Picasso? That's out of question today and generative adversarial networks (GANs \cite{GAN}) can help you producing images which look like paintings by the artist you choose.

GAN is a new framework of training generative models and has drawn significant attention in recent years.
The idea of GANs is to introduce a discriminator to play the role of the generator's adversary, so their competition forms a two-player game. The objective of the discriminator is to distinguish real samples from generated ones, \emph{e.g.}, fake images, while the objective of the generator is to produce new instances resembling training instances, \emph{e.g.}, real images.

Compared with other generative models with explicit likelihoods such as Boltzmann machines, belief networks and autoencoders, GAN as an implicit generative model can be more expressive due to less restrictions in network structures. Many variations of GAN have been proposed including conditional GAN \cite{MS14}, LAPGAN \cite{DCF15}, DCGAN \cite{RMC15} and InfoGAN \cite{CDH16}.

Apart from images, language data also plays an important role in artificial intelligence. As the great ideologist Denis Diderot said, ``If they find a parrot who could answer to everything, I would claim it to be an intelligent being without hesitation \ldots''
You may wonder whether GANs can help with such dialogue generators, or generate other language and text data, \emph{e.g.}, poetry, stories and jokes. Indeed, GANs could potentially become powerful tools for natural language processing, but the first condition is that the vanishing gradient problem of training GANs can be resolved, since language data is much more discrete than data in the visual domain.

\begin{figure}
\centering
\subfigure[update the parameters of $D$]{
\includegraphics[height=6cm]{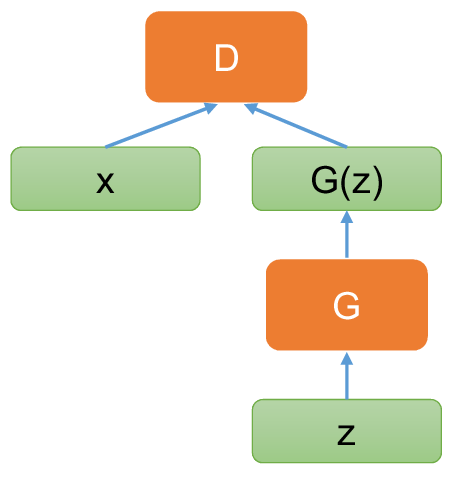}}
\hspace{2cm}
\subfigure[update the parameters of $G$]{
\includegraphics[height=6cm]{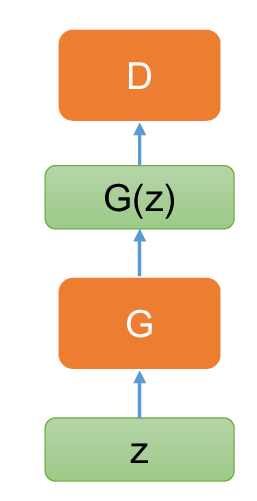}}
\caption{Training process of GAN. Step (a) and (b) are alternately carried out until the generator $G$ outputs the same data distribution as the target distribution.}
\label{fig:GAN}
\end{figure}

Before we explain the vanishing gradient problem of training GANs, let's have a closer look at the training process.
The process of training the discriminator $D$ and the generator $G$ alternately is depicted in Fig. \ref{fig:GAN}.
In step (a), $G$ receives a random noise $z$ and then produces some fake samples $G(z)$, then $D$ is trained to distinguish between training data $x$ and generated data $G(z)$. The parameters of $D$ are updated in order to maximize
\begin{align}
\mathbb{E}_{x\sim P_{d}(x)} \log D(x) + \mathbb{E}_{z\sim P_{z}(z)} \log \big(1-D(G(z))\big),
\end{align}
where $P_d(x)$ is the real data distribution and $P_{z}(z)$ is the distribution of the input noise. The output $D(x)\in [0,1]$ can be explained as the probability that $D$ thinks the sample $x$ is real.
The purpose of this step is to make $D$ a better adversary, so $G$ has to try harder to fool $D$.
In step (b), the output samples of the generator are labeled as real and then fed into $D$.
The parameters of $G$ are updated in order to maximize
\begin{align}
\mathbb{E}_{z\sim P_{z}(z)} \log D(G(z)),
\end{align}
trying to make $D$ believe the generated samples $G(z)$ are real.
After this step, the ability of $G$ improves a little bit.
By repeating these two steps, the game will reach an equilibrium point, where the generator is able to generate the same statistics as the real samples and the
prediction accuracy of the discriminator is 1/2, not better than a random guess.

The vanishing gradient problem occurs in step (b).
If the output $G(z)$ is continuous with respect to the variation of each parameter of $G$, the mapping $D(G(z))$ is a continuous function.
In contrast, if the output $G(z)$ is discrete with respect to the variation of each parameter of $G$, the mapping $D(G(z))$ is a staircase function, so the gradient of $D(G(z))$ with respect to each parameter of $G$ is zero almost everywhere. The training fails because the parameters cannot be updated according to zero gradients. Therefore, it is crucial to address  the vanishing gradient problem  before GAN being capable of generating discrete data.

The era of quantum computing is around the corner. In 2016, IBM provided access to its quantum computer to the community through a cloud platform called IBM Quantum Experience \cite{IBM}. A quantum computing competition among IT giants including Microsoft, Google, Intel is under way.
Quantum computing is a new computing model in which computing tasks are accomplished by controlling quantum information units under the rules of quantum mechanics.
Because quantum computing has the admirable capability of processing exponentially high-dimensional data,
quantum machine learning \cite{BWP17,NSR} is expected to be one of the most intriguing future applications of quantum computers.
Quantum machine learning is the crossover between machine learning and quantum computing which goes both ways. On one hand, machine learning algorithms can be used to improve the benchmarking, understanding and controlling of quantum computing systems. On the other hand, quantum computing promises significant enhancement of some machine learning algorithms, especially in computational complexity.
Many theoretical and experimental researches \cite{HHL,WBL12,CWS13,QSVM,QPCA,CWS15,YGW16,DTB16,MSW17,DYL17,YGL18, NM00, CCW15} on machines learning problems with the help of quantum computing have been taken in the last decade. However, most of the existing quantum  learning algorithms are confronted with the input/output bottlenecks. More specifically,  some quantum algorithms  require that the classical input data must be encoded into  quantum states in advance, which thus consumes much additional time and inevitably compromises the speedup of the quantum algorithm. Other algorithms only output a quantum state corresponding to the solution, instead of directly giving the solution itself. Thus,  extra time is required if one wants to extract information from the output quantum states. Consequently, how to avoid these problems is of great importance for a quantum learning algorithm to be used in practice.

\subsection{Our contribution}
In this paper, by observing the vanishing gradient problem of classical GAN when generating discrete data, we explore the ability of GAN based on quantum computing which naturally has merit to equip GANs with the ability of dealing with discrete data. Meanwhile, in order to avoid the input/output bottlenecks embarrassing   current quantum learning algorithms,  our model is designed to be a classical-quantum hybrid architecture which receives its input in the classical form by the classical discriminator and  outputs the discrete data by performing a measurement on the state generated from a quantum circuit. More specifically, we propose a quantum GAN for generating classical discrete distribution, which is composed of a parameterized quantum circuit as the generator and a classical feedforward neural network as the discriminator. The parameterized quantum circuit only consists of simple one-qubit rotation gates and two-qubit controlled-phase gates that are available in current quantum devices. We present a small-scale numerical simulation to demonstrate the  effectiveness of our scheme.

Our scheme shows characteristics and potential advantages as follows: (i) It can generate discrete data, thus complements classical GANs that are  clumsy for this task due to the problem of vanishing gradient. (ii) It avoids the input/output problem as stated before, since both the input and the output are directly given in the classical form, which makes our scheme more feasible in practice, compared with other quantum  learning algorithms. (iii) The probability distribution  implicitly given by data samples  can be loaded into a quantum state, which may be useful for some further applications.

\subsection{Related work}

Recently the mergence of machine learning and quantum computing has developed into a hot topic, and some interesting results have been obtained \cite{BWP17,NSR}.
Therefore, it is natural to consider how to improve the generative models with the help of quantum computing. Actually, some efforts have been devoted to this issue. For example, Ref. \cite{BGN18} trained shallow parameterized quantum circuits to generate GHZ states, coherent thermal states and Bars and Stripes images.
Ref. \cite{LiuWang18} developed a gradient-based learning scheme to train deep parameterized quantum circuits for generation of Bars and Stripes images and mixture of Gaussian distributions. These quantum generative models are also known as Born machines as the output probabilities are determined by Born's rule \cite{Born}. In addition,
the idea of quantum generative adversarial learning was recently explored theoretically in Ref. \cite{LW18}. A quantum GAN consists of a quantum generator and a quantum discriminator was numerically implemented to generate simple quantum states \cite{DK18}.
Ref. \cite{BGW18} derived an adversarial algorithm for the problem of approximating an unknown quantum pure state.
Ref. \cite{HWC18} demonstrated that a superconducting quantum circuit can be adversarially trained to replicate the statistics of the quantum data output from a digital qubit channel simulator. Compared with these researches on quantum GANs \cite{LW18,DK18,BGW18,HWC18} that focused on generating quantum data, our work centers on the generation of classical discrete data.

Parameterized quantum circuits are also used in other machine learning parameterized models \cite{WDK17,ROA17,MNK18,CSS18,LAM19}.  One of the possible reasons for adopting parameterized quantum circuits is that sampling from output distributions of random quantum circuits must take exponential time in a classical computer \cite{BIS18}, which suggests that quantum circuits exhibit stronger representational power than neural networks.

Besides GANs, there is another commonly used generative model, called variational autoencoders (VAE \cite{kingma2013auto}), which can also be improved with quantum techniques. Ref. \cite{khoshaman2018quantum} introduced quantum VAEs and used quantum Monte Carlo simulations to train and evaluate the performance of quantum VAEs.


Note that the original version of this paper is given in \cite{Situ18}. After our work was completed, we found that some researchers were also interested in this topic and wrote a similar manuscript \cite{ZWL18}.  However, our work includes a general quantum circuit and an MPS quantum circuit as the architecture of the generator, while they only use the general quantum circuit. In addition, we report the averages and standard deviations of 30 experiments to show the reliable performance of our scheme, while they only report the result of a single experiment.
Also note that after our work, more attention has been paid to quantum GANs \cite{CHL19,ZLW19,BLS19,RA19,DHT19}.

\subsection{Nomenclatures}
We first introduce some nomenclatures for readers who are not familiar with quantum computing. For more concrete explanations about these concepts, please refer to the textbook \cite{textbook}.
\textit{Qubits} (quantum bits) are the fundamental objects of information in quantum computing. In classical computing, a bit can have value either 0 or 1. However, a qubit can be a superposition of $|0\rangle$ and $|1\rangle$. The state of a single qubit can be described by a two-dimensional complex valued column vector of unit norm. For example, qubits $|0\rangle$ and $|1\rangle$ can be described by $[1\ 0]^T$ and $[0\ 1]^T$. The transposed-conjugates of $|0\rangle$ and $|1\rangle$ are denoted as $\langle 0|$ and $\langle  1|$.
\textit{Quantum state (wave function)} is the state of an isolated quantum system. An $N-$qubit quantum state (wave function) is described by a $2^N-$dimensional complex valued column vector of unit norm.
\textit{Quantum gate} is a physical operation that acts on qubits. A quantum gate on $N$ qubits can be described by a $2^N \times 2^N$ unitary matrix.
\textit{Quantum circuit} is a sequence of quantum gates, which are reversible transformations on a quantum mechanical analog of an $N-$bit register.
\textit{Quantum measurement} is a readout process of quantum information. Unlike classical information, a quantum state may change after being measured.

\subsection{Organization}
This paper is organized as follows. In section \ref{sec:model}, we present the constituents of the quantum generator and the classical discriminator. The loss function and optimization method are also described. Then the adversarial training algorithm is provided, together with the gradient estimation method for updating the parameters of the quantum generator.
In section \ref{sec:experiment}, we report the numerical simulation  testifying the effectiveness of our scheme.
In section \ref{sec:discussion}, we discuss our scheme and the future work.
A brief conclusion follows in section \ref{sec:conclusion}.

\section{Model architecture and training method}\label{sec:model}
In this section, we present the architecture of our generative quantum circuits built with simple one-qubit rotation and two-qubit controlled-phase gates, and the adversarial training scheme.

\subsection{Generative quantum circuit}\label{sec:generator1}
\begin{figure}
\centering
\includegraphics[width=4in]{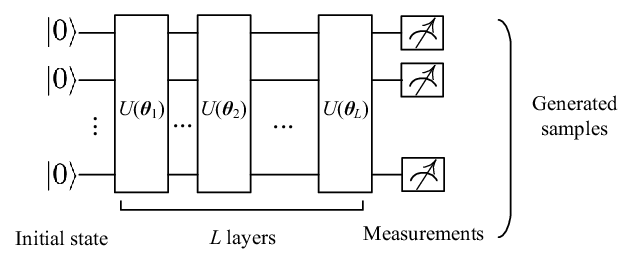}
\caption{The generative quantum circuit with $L$ layers}
\label{fig:QuantGen}
\end{figure}

\begin{figure}
\centering
\includegraphics[width=5in]{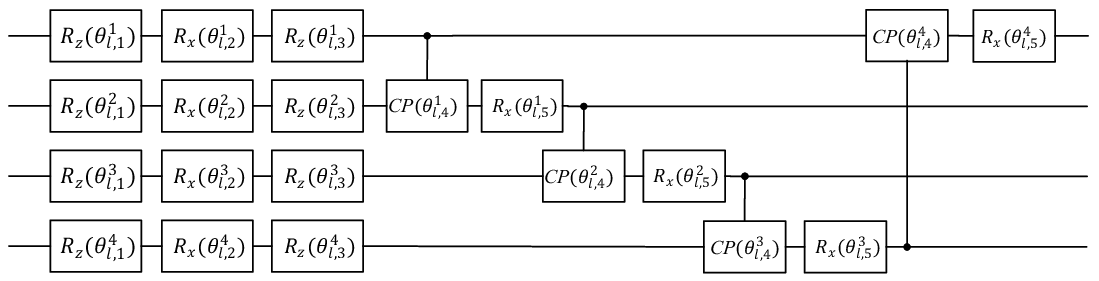}
\caption{A layer of the quantum circuit for four qubits}
\label{fig:ansatz}
\end{figure}

Our quantum circuit for generation of $N$-bit samples involves $N$ qubits, the layout of which is described in Fig. \ref{fig:QuantGen}.
The input quantum state is initialized to $|0\rangle^{\otimes N}$, and then passed through $L$ layers of unitary operations.
At the end of the circuit, all the qubits are measured in the computational basis.
The measurement outcomes are gathered to form an $N$-bit sample $x$.
Each layer is composed of several one-qubit rotation gates and two-qubit controlled-phase gates.
Fig. \ref{fig:ansatz} shows the arrangement of these gates in one layer.
Three rotation operations are first applied to each qubit. This process can be written as
\begin{align}
\prod_{i=1}^{N}R_z^{i}(\theta^{i}_{l,3})R_x^{i}(\theta^{i}_{l,2})R_z^{i}(\theta^{i}_{l,1}),
\end{align}
where the superscript $i$ denotes the $i$th qubit, and the subscript $l$ denotes the $l$th layer.
$R_x(\theta)$ and $R_z(\theta)$ are rotation gates, \emph{i.e.},
\begin{align}
R_x(\theta)=\left(
\begin{array}{cc}
  \cos\frac{\theta}{2} & -i\sin\frac{\theta}{2} \\
  -i\sin\frac{\theta}{2} & \cos\frac{\theta}{2} \\
\end{array}
\right),
R_z(\theta)=\left(
\begin{array}{cc}
  e^{-i\theta/2} & 0 \\
  0 & e^{i\theta/2} \\
\end{array}
\right).
\end{align}
The number of parameters/gates in this process is $3N$ per layer.
The choice of these operators is because any one-qubit unitary can be decomposed  into this sequence of rotation operators \cite{textbook}.

We also need to entangle the qubits by performing controlled-$U$ gates between the qubits. This process can be written as
\begin{align}
\prod_{i=1}^{N}CU^i_{(i\ \mathrm{mod}\ N)+1},
\end{align}
where the superscript $i$ denotes the control qubit, and the subscript $(i\ \mathrm{mod}\ N)+1$ denotes the target qubit.
Each unitary is characterized by three parameters, so the number of parameters in this process is $3N$ per layer.
However, Ref. \cite{SBS18} has pointed out that this process can be simplified to
\begin{align}
\prod_{i=1}^{N}R_x^{(i\ \mathrm{mod}\ N)+1}(\theta^i_{l,5})CP^i_{(i\ \mathrm{mod}\ N)+1}(\theta^i_{l,4}),
\end{align}
where
\begin{align}
CP(\theta) = \left(
\begin{array}{cccc}
  1 & 0 & 0 & 0 \\
  0 & 1 & 0 & 0 \\
  0 & 0 & 1 & 0 \\
  0 & 0 & 0 & e^{i\theta} \\
\end{array}
\right)
\end{align}
is the controlled-phase gate.
Now the entangling process only has $2N$ parameters/gates per layer. The total number of parameters/gates in the quantum circuit is $5NL$.
The set of all parameters can be denoted as a vector $\vec{\theta}=\{\theta_1,\ldots,\theta_{5NL}\}$ for convenience of expression.

\subsection{Generative MPS quantum circuit}\label{sec:generator2}

\begin{figure}
\centering
\includegraphics[width=3in]{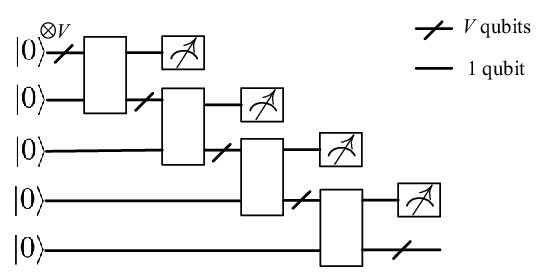}
\caption{The generative MPS quantum circuit with $N=4$ nodes}
\label{fig:MPS}
\end{figure}

There are limited number of qubits in near-term quantum devices. However, many real data sets consist of millions of samples and thousands of features. Thus, it is crucial to design quantum algorithms which require less qubits.
Besides the aforementioned family of quantum circuits, we also consider another family of quantum circuits,
which are called ``Matrix Product State (MPS) quantum circuits'' \cite{HPW18}.
The quantum generator can be implemented by MPS quantum circuits with less qubits.
Fig. \ref{fig:MPS} illustrates the structure of the MPS quantum circuit,
which looks like a maximally unbalanced tree with $N$ nodes.
Each node is a quantum ansatz which inputs and outputs $V+1$ qubits.
The uppermost output qubit of each node is measured in the computational basis and the other $V$ qubits flow to the next node.
The $N$ measurement outcomes comprise the $N$-bit generated sample $x$.
Each node can contain $L\geqslant 1$ layers which have the same gates and layouts as the layers depicted in Fig. \ref{fig:ansatz}.
The number of parameters/gates in one node is $5L(V+1)$, so the number of parameters/gates in an MPS quantum circuit is $5NL(V+1)$.
The input qubits are all initialized to $|0\rangle$ in our numerical experiment.

\begin{figure}
\centering
\includegraphics[width=5in]{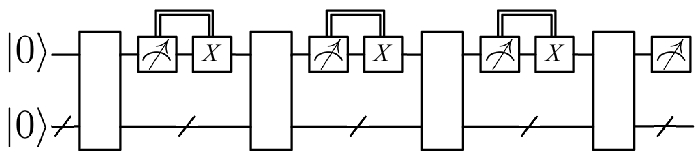}
\caption{The generative MPS quantum circuit with reused qubits}
\label{fig:MPSreuse}
\end{figure}

If the MPS quantum circuit is implemented using quantum devices,
each qubit that has been measured can be set to $|0\rangle$ and reused as the input of the next node.
So only $V+1$ qubits are actually needed in the circuit evaluation process.
The sample dimension $N$ is only related to the depth of the circuit.
Fig. \ref{fig:MPSreuse} gives an equivalent form of the MPS circuit in order to illustrate the idea of qubit recycling. Double line represents a measurement result which is a classical bit 0 or 1. This bit is  used as a control signal of a controlled-$X$ (i.e., controlled-NOT) gate. That is, the $X$ gate (i.e., NOT gate) is performed only if this bit is 1. If the measurement result is 0, the qubit after measurement is in the $|0\rangle$ state. If the measurement result is 1, the qubit after measurement is in the $|1\rangle$ state, and subsequently flipped to $|0\rangle$ by the $X$ gate. In this way, the qubit being measured is reset to $|0\rangle$. This quantum circuit has advantage in physical implementation because near-term quantum devices have limited number of qubits.

\subsection{Discriminator}

A discriminator $D$ is introduced to distinguish between real samples and generated samples. Recently, quantum-classical hybrid architectures are often adopted in  quantum  algorithms due to the limitations of the coherence time and qubit number in near-term quantum device. The discriminative task in quantum GANs for discrete data generation is a simple two-class classification problem in which traditional feedforward neural network is efficient and has satisfying performance.
The input layer of our discriminator has the same dimension as the samples. Only one hidden layer is employed.
The output layer has only one output value in $[0,1]$, which represents the discriminator's prediction about the probability of the input sample being real.
An output $D(x)=1$ means the discriminator believes the input sample $x$ is definitely real,
while an output $D(x)=0$ means it believes the input sample $x$ is definitely fake.

The loss function of the discriminator we adopt here is the binary cross entropy function commonly used in binary classification tasks:
\begin{align}\label{eq:BCE-exp}
J_D = -\frac{1}{2}\Big(\mathbb{E}_{x\sim P_{d}(x)} \log D(x) + \mathbb{E}_{x\sim P_{\vec{\theta}}(x)} \log \big(1-D(x)\big)\Big),
\end{align}
where $P_d(x)$ is the real data distribution and $P_{\vec{\theta}}(x)$ is the generated distribution.
$\mathbb{E}$ denotes the expectation operator. As the distributions $P_d(x)$ and $P_{\vec{\theta}}(x)$ are typically unknown, $J_D$ can be empirically estimated from a set of available samples drawn from $P_d(x)$ and $P_{\vec{\theta}}(x)$. In every epoch of the training process, we sample one mini-batch of samples from the real data and the generator, respectively. Then the average loss can be calculated by
\begin{align}\label{eq:BCE}
J_D(\mathbf{x},\mathbf{y}) = -\frac{1}{2\cdot\mathbf{batch\_D}}\sum_i y_i \log D(x_i) + (1-y_i) \log (1-D(x_i)),
\end{align}
where $\mathbf{batch\_D}$ denotes the number of samples in one mini-batch.
The loss function evaluates how close are the predictions $D(x_i)$ and the desired labels $y_i$.
$J_D(\mathbf{x},\mathbf{y})$ achieves minimum zero if $D(x_i)=y_i$ for every $(x_i,y_i)$.

Let $\mathbf{w}$ be the set of all the parameters of the discriminator.
The gradient of $J_D(\mathbf{x},\mathbf{y})$ with respect to $\mathbf{w}$ can be obtained by the backpropagation algorithm. A variety of gradient-based optimization algorithms can be used to train the discriminator.
For example, the vanilla gradient descent method updates $\mathbf{w}$ in the following way:
\begin{align}
\mathbf{w} \leftarrow \mathbf{w} - \alpha_D \cdot \frac{\partial J_D(\mathbf{x},\mathbf{y})}{\partial \mathbf{w}},
\end{align}
where $\alpha_D$ is the learning rate, $\partial$ denotes partial derivative.

\subsection{Optimization of the generator parameters}

The goal of the generator is to generate samples that can fool the discriminator.
The training process of the generator only uses generated samples, which are paired with true labels, so Eq. (\ref{eq:BCE-exp}) reduces to
\begin{align}
J_G = -\mathbb{E}_{x\sim P_{\vec{\theta}}(x)} \log D(x),
\end{align}
where $P_{\vec{\theta}}(x)$ is the probability of getting measurement outcome $x$ from the quantum circuit parameterized with $\vec{\theta}=\{\theta_1,\theta_2,\ldots\}$.
The gradient of $J_G$ with respect to a single parameter $\theta\in\vec{\theta}$ is
\begin{align}\label{eq:gradient1}
\frac{\partial J_G}{\partial \theta} = - \sum _{x\in\{0,1\}^N} \log D(x)\frac{\partial P_{\vec{\theta}}(x)}{\partial \theta}.
\end{align}

The following formulas are needed in the calculation of $\frac{\partial P_{\vec{\theta}}(x)}{\partial \theta}$, which can be proved directly by using the definitions of matrix multiplication and tensor product (denoted as $\otimes$).
\begin{align}
& \frac{\partial ABC}{\partial \theta}=A\frac{\partial B}{\partial \theta}C,\\
& \frac{\partial A\otimes B\otimes C}{\partial \theta}=A\otimes \frac{\partial B}{\partial \theta}\otimes C,
\end{align}
where the parameter $\theta$ appears in operator $B$ but not in $A$ or $C$. Another useful formula is
\begin{align}
\frac{\partial AB}{\partial \theta}=\frac{\partial A}{\partial \theta}B+A\frac{\partial B}{\partial \theta},
\end{align}
where the parameter $\theta$ appears in both operators $A$ and $B$.

Suppose the quantum circuit has $k$ gates which can be denoted as $U_j$ with $j\in{\{1,\ldots,k\}}$. For an initial state $\rho_0$, the output of the quantum circuit is $U_{k:1}\rho_0 U^\dagger_{k:1}$, where the notation $U_{k:1}=U_k\ldots U_1$ is introduced for convenience, and $\dagger$ means transposed-conjugate.
Suppose $\theta$ is a parameter that appears and only appears in $U_j$, the partial derivative of $P_{\vec{\theta}}(x)$ with respect to $\theta$ is
\begin{align}\label{eq:A1}
\frac{\partial P_{\vec{\theta}}(x)}{\partial \theta} & =  \frac{\partial}{\partial \theta} \langle x|U_{k:1}\rho_0 U^\dagger_{k:1}|x\rangle \nonumber\\
& = \langle x| U_{k:j+1} \frac{\partial U_j}{\partial \theta} U_{j-1:1}\rho_0 U^\dagger_{k:1}|x\rangle + \langle x| U_{k:1} \rho_0 U^\dagger_{j-1:1} \frac{\partial U^\dagger_j}{\partial \theta}  U^\dagger_{k:j+1}|x\rangle.
\end{align}
For $U_j$ and $U^\dagger_j$ being a controlled-phase gate and its Hermitian conjugate, the gradients are
$\frac{\partial U_j}{\partial \theta}=i|11\rangle\langle 11| U_j$ and
$\frac{\partial U^\dagger_j}{\partial \theta}=-i U^\dagger_j|11\rangle\langle 11|$, respectively. By substituting them into Eq. (\ref{eq:A1}), we have
\begin{align}\label{eq:A2}
\frac{\partial P_{\vec{\theta}}(x)}{\partial \theta} & = i \langle x| U_{k:j+1} |11\rangle\langle 11| U_{j:1} \rho_0 U^\dagger_{k:1}|x\rangle
- i \langle x| U_{k:1} \rho_0 U^\dagger_{j:1} |11\rangle\langle 11| U^\dagger_{k:j+1} |x\rangle\nonumber\\
& = i\langle x| U_{k:j+1} [|11\rangle\langle 11|,U_{j:1}\rho_0 U^\dagger_{j:1}] U^\dagger_{k:j+1} |x\rangle.
\end{align}
The following property of the commutator for an arbitrary operator $\rho$
\begin{align}\label{eq:A3}
[|11\rangle\langle 11|,\rho] = -\frac{i}{2} \Big(U_j(\frac{\pi}{2})\rho U^\dagger_j(\frac{\pi}{2}) - U_j(-\frac{\pi}{2})\rho U^\dagger_j(-\frac{\pi}{2})\Big)
\end{align}
can be verified by the substitution of $U_j(\frac{\pi}{2})=I+(i-1)|11\rangle\langle 11|$ and
$U_j(-\frac{\pi}{2})=I-(1+i)|11\rangle\langle 11|$.
By substituting Eq. (\ref{eq:A3}) into Eq. (\ref{eq:A2}), we have

\begin{align}
\frac{\partial P_{\vec{\theta}}(x)}{\partial \theta}  = & \frac{1}{2} \Big(\langle x|U_{k:j+1} U_j(\theta+\frac{\pi}{2})U_{j-1:1}\rho_0 U^\dagger_{j-1:1} U^\dagger_j(\theta+\frac{\pi}{2}) U^\dagger_{k:j+1}|x\rangle \nonumber\\
& -  \langle x|U_{k:j+1} U_j(\theta-\frac{\pi}{2})U_{j-1:1}\rho_0 U^\dagger_{j-1:1} U^\dagger_j(\theta-\frac{\pi}{2}) U^\dagger_{k:j+1}|x\rangle\Big)\nonumber\\
= & \frac{1}{2}(P_{\vec{\theta}^{+}}(x)-P_{\vec{\theta}^{-}}(x)),
\end{align}
where $\vec{\theta}^{\pm} = \vec{\theta} \pm \frac{\pi}{2} \mathbf{e}^i$, $\mathbf{e}^i$ is the $i$th unit vector in the parameter space corresponding to $\theta$
(i.e., $\theta \leftarrow \theta \pm \frac{\pi}{2}$, with other angles $\vec{\theta}-\{\theta\}$ unchanged).
For $U_j$ being the rotation gates $R_x,R_y$ or $R_z$, Ref. \cite{MNK18} has proved the similar equation
\begin{align}\label{eq:gradient2}
\frac{\partial P_{\vec{\theta}}(x)}{\partial \theta} =\frac{1}{2}\big(P_{\vec{\theta}^{+}}(x)-P_{\vec{\theta}^{-}}(x)\big).
\end{align}

By substituting Eq. (\ref{eq:gradient2}) into Eq. (\ref{eq:gradient1}), we have
\begin{align}
\frac{\partial J_G}{\partial \theta} = \frac{1}{2} \mathbb{E}_{x\sim P_{\vec{\theta}^-}(x)}\log D(x) - \frac{1}{2} \mathbb{E}_{x\sim P_{\vec{\theta}^+}(x)}\log D(x).
\end{align}
In order to estimate the gradient with respect to each $\theta\in \vec{\theta}$,
we have to sample two mini-batches $\mathbf{x^+}$ and $\mathbf{x^-}$ from the circuits with parameters $\vec{\theta}^+$ and $\vec{\theta}^-$, respectively, then the gradient is estimated by
\begin{align}\label{eq:gEst}
 \frac{1}{2\cdot \mathbf{batch\_G}} \Big(\sum_{x\in\mathbf{x^-}}\log D(x) -  \sum_{x\in\mathbf{x^+}}\log D(x)\Big),
\end{align}
where $\mathbf{batch\_G}$ denotes the number of samples in one mini-batch.

The generator's parameters $\vec{\theta}$ can be optimized by gradient-based optimization algorithms. For example, the vanilla gradient descent method updates $\vec{\theta}$ in the following way:
\begin{align}\label{eq:updateG}
\vec{\theta} \leftarrow \vec{\theta} - \alpha_G \cdot \frac{\partial J_G}{\partial \vec{\theta}},
\end{align}
where $\alpha_G$ is the learning rate.

It is worth pointing out that in the classical case, $\frac{\partial J_G}{\partial \vec{\theta}}$  in  Eq. (\ref{eq:updateG}) will be zero (that is,  the gradient vanishes), when GANs are trained to generate discrete data. However, in our quantum GAN,  $\frac{\partial J_G}{\partial \vec{\theta}}$ can be obtained by   Eq. (\ref{eq:gradient2}), which has no counterpart in the classical scenario.
Note that classical generators map a random distribution to the target distribution, while the randomness of the quantum circuit output stems from the final measurement. So quantum GAN is intrinsically capable of generating discrete data, while classical GANs are clumsy for this task due to the vanishing gradient problem. In addition, our quantum GAN has a classical-quantum hybrid architecture which receives its input in the classical form by the classical discriminator and  outputs the discrete data by performing a measurement on the state generated from the quantum circuit. Thus, it will not encounter the input/output bottlenecks embarrassing most of the existing quantum learning algorithms. Finally, as one can see, when the adversarial training procedure is completed, the probability distribution  implicitly given by data samples will be loaded into a quantum state, which can be explicitly calculated by performing the obtained quantum circuit to the initial quantum state. However, this is impossible in the classical case.  This property of quantum GAN may be useful for some further applications.

\subsection{Adversarial training}

The adversarial training algorithm of our quantum GAN is described in Algorithm \ref{alg:QGAN}.
The training process iterates for a fixed number of epoches, or until some stopping criterion is reached, \emph{e.g.}, convergence on the loss function. At each epoch, the parameters of the discriminator and the generator are updated $\mathbf{d\_step}$ and $\mathbf{g\_step}$ times, respectively.

\begin{algorithm}[t]
\renewcommand{\algorithmicrequire}{\textbf{Input:}}
\renewcommand\algorithmicensure {\textbf{Output:} }
\caption{Adversarial training algorithm of our quantum GAN}
\label{alg:QGAN}
\begin{algorithmic}[1]
\Require $L$: number of layers; $V$: number of ancilla qubits (only for MPS circuits); $\mathbf{batch\_D},\mathbf{batch\_G}$: mini-batch size; $\mathbf{d\_step}$: times of updating $\mathbf{w}$ in one epoch; $\mathbf{g\_step}$: times of updating $\vec{\theta}$ in one epoch;

\Ensure $\vec{\theta}$: the parameters of the generator
\State Initialize the generator and the discriminator with random parameters
\For{number of training epoches}
    \For{$\mathbf{d\_step}$ steps}
        \State Sample a mini-batch of  $\mathbf{batch\_D}$ samples from the training dataset. Label them as ``real''.
        \State Sample a mini-batch of $\mathbf{batch\_D}$ samples from the quantum circuit. Label them as ``fake''.
        \State Use these samples and labels to calculate the gradient of the loss according to Eq. (\ref{eq:BCE}).
        \State Update the discriminator's parameters $\mathbf{w}$ according to the gradient.
    \EndFor

    \For{$\mathbf{g\_step}$ steps}
        \State For each $\theta_i$, sample a mini-batch of $\mathbf{batch\_G}$ samples from the quantum circuit with parameters $\vec{\theta}^{+}$.
        \State For each $\theta_i$, sample a mini-batch of $\mathbf{batch\_G}$ samples from the quantum circuit with parameters $\vec{\theta}^{-}$.
        \State Use these samples to calculate the gradient of the loss according to Eq. (\ref{eq:gEst}).
        \State Update the generator's parameters $\vec{\theta}$ according to the gradient.
    \EndFor
\EndFor
\end{algorithmic}
\end{algorithm}

\section{Numerical simulation}\label{sec:experiment}

We verify our proposal using a synthetic dataset known as Bars and Stripes (BAS),
which is also used in Ref. \cite{BGN18,LiuWang18} to test quantum generative models.
The dataset contains $m\times m$ binary images with only bar patterns or stripe patterns.
There are $2^m$  different vertical bar patterns and $2^m$ different horizontal stripe patterns.
The all-black and all-white patterns are counted in both bar patterns and stripe patterns.
So there are $2^{m+1}-2$ possible BAS patterns for an $m\times m$ image.
We assume all BAS patterns appear with equal probability.
Obviously each pixel can be encoded in one qubit, so an $m\times m$ image can be encoded in $m^2$ qubits.
We restrict our experiments to the case of $m=2$, because the simulation time on classical computers grows exponentially with the number of qubits. For instance, in the case of $m=3$, a single run of 5000 epoches  costs more than a week on an ordinary PC (Intel Core i7-6700 3.40 GHz CPU, 8.00 GB RAM), in order to produce a decent distribution. However, the rapid increment of computation complexity does not happen in the real quantum device which ensures the proposed quantum GAN can be applied to generate real-world data, e.g., poem.
 Experiments for larger $m$ and some real-world dataset may be done in the future if an intermediate-scale near-term quantum device is available.

The simulation code is written in Python language.
The discriminator is classical so it's implemented using the widely used deep learning framework PyTorch \cite{PyTorch}.
The discriminator has one input layer with dimension $m \times m$, one hidden layer made up of 50 neurons with the ReLU activation function $f(x)=\max(0,x)$,
and one output neuron using the Sigmoid activation function $f(x)=1/(1+e^{-x})$.
Learning rate is a hyper-parameter that controls how much we adjust the parameters in each epoch. A larger learning rate results in a faster learning speed, but the found solution may be oscillating around the optimal solution. A small learning rate can ensure the convergence to the optimal solution but the learning process may be too slow. Dozens of advanced gradient descent methods have been proposed to help with fast convergence to the optimal solution. We adopt the stochastic gradient optimizer Adam (Adaptive Moment Estimation) \cite{Adam} to update the discriminator's parameters with the suggested initial learning rate $10^{-3}$.

The generative quantum circuit is simulated directly by calculating the evolution of the wavefunction.
An $N$-qubit wavefunction is encoded in a $2^N$-dimensional complex vector.
After performing a single-qubit operation $u_{11}|0\rangle\langle0|+u_{12}|0\rangle\langle1|+u_{21}|1\rangle\langle0|+u_{22}|1\rangle\langle1|$ on the $i$th qubit, the wavefunction is transformed to
\begin{align}
\alpha'_{*\ldots *0_i*\ldots *} = u_{11}\cdot \alpha_{*\ldots *0_i*\ldots *} + u_{12}\cdot \alpha_{*\ldots *1_i*\ldots *},\\
\alpha'_{*\ldots *1_i*\ldots *} = u_{21}\cdot \alpha_{*\ldots *0_i*\ldots *} + u_{22}\cdot \alpha_{*\ldots *1_i*\ldots *},
\end{align}
where $\alpha$ and $\alpha'$ are amplitudes before and after transformation.
The case of two-qubit operation can be deduced analogously.
The learnable parameters of the quantum circuit are radians, which are randomly initialized in the interval $(-\pi,\pi)$.
They are updated without constraint according to Eq. (\ref{eq:updateG}) with a constant learning rate $\alpha_G=2\times 10^{-2}$.
The gradient is estimated according to Eq. (\ref{eq:gEst}).

In general, a larger mini-batch size provides a computationally more efficient process while a smaller size allows for a more robust convergence and avoiding local minima. We find that the mini-batch sizes have little influence on our experiment results. The mini-batch sizes in the following experiments we reported are $\mathbf{batch\_D}=64$, $\mathbf{batch\_G}=100$.

In each epoch, the discriminator's parameters are updated $\mathbf{d\_step}$ times, and then the generator's parameters are updated $\mathbf{g\_step}$ times. The default values of $\mathbf{d\_step}$ and $\mathbf{g\_step}$ are both 1. We also adopt these default values in our experiments.

The two numerical experiments we perform differ in the structure of the quantum generator.
The first experiment uses the general quantum circuit described in section \ref{sec:generator1},
while the second experiment uses the MPS quantum circuit described in section \ref{sec:generator2}.

\subsection{Numerical experiment 1}

In the first numerical experiment, the quantum generator is the general quantum circuit presented in section \ref{sec:generator1}.
Unlike the training of classification models, the stopping criterion of training GANs is very tricky, so we simply run the training algorithm for 5000 epoches.
For each $L=1,2,3,4,5,6$, we repeat the experiment 30 times. The averages and standard deviations of three indicators (i.e., accuracy, KL divergence and loss) are reported every 50 epoches.

\begin{figure}
\centering
\includegraphics[width=4in]{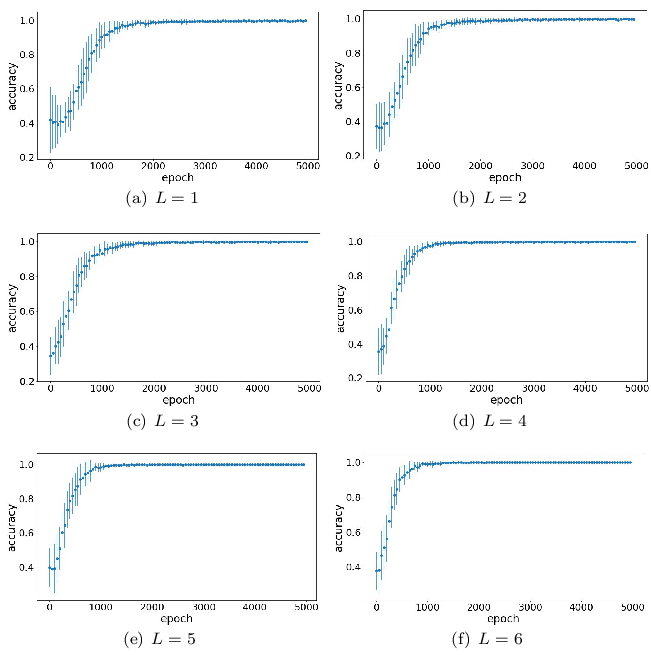}
\caption{Averages and standard deviations of the accuracy w.r.t. the number of epoches}
\label{fig:accuracy1}
\end{figure}

We first examine the accuracy of the generator.
The accuracy in some epoch is defined as the ratio of the number of valid samples (i.e., BAS patterns) in one mini-batch  to $\mathbf{batch\_D}$.
The generator accuracy w.r.t. the number of epoches is depicted in Fig. \ref{fig:accuracy1}.
We can see that for each $L$ from 1 to 6, the accuracy increases very quickly and achieves nearly $100\%$ in 1000 epoches,
which means that it's not difficult for the generator to learn to avoid producing non-BAS patterns.

\begin{figure}
\centering
\includegraphics[width=4in]{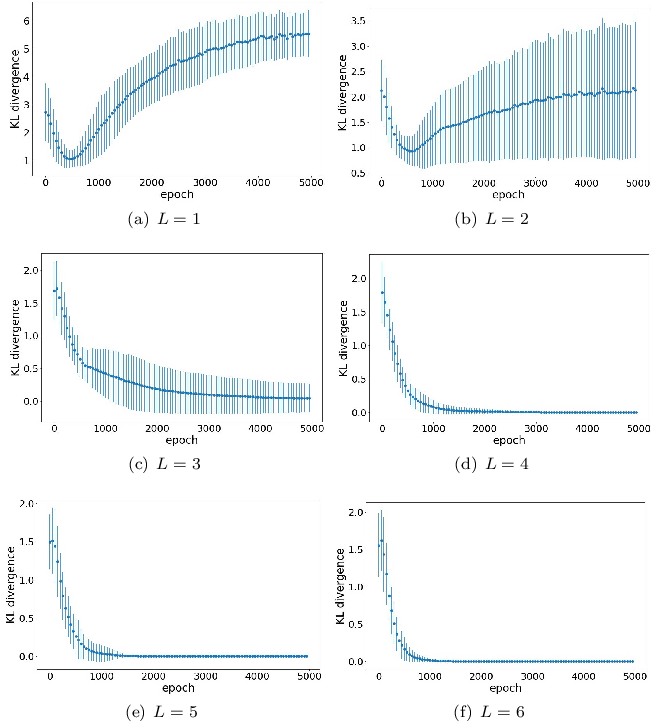}
\caption{Averages and standard deviations of the KL divergence w.r.t. the number of epoches}
\label{fig:KLD1}
\end{figure}

But our goal is not merely producing correct BAS patterns.
The distribution of the generated patterns is expected to be the same as that of the training dataset, i.e., uniform distribution in our case.
KL divergence is usually used to measure how one probability distribution diverges from a second, expected probability distribution, which is defined by
\begin{align}
\mathrm{KLD}(P_d\|P_{\vec{\theta}}) = - \sum_x P_d(x) \log\frac{P_{\vec{\theta}}(x)}{P_d(x)},
\end{align}
where $P_d$ and $P_{\vec{\theta}}$ are the real data distribution and the generated distribution, respectively.
$\mathrm{KLD}(P_d\|P_{\vec{\theta}})$ is non-negative and equals zero if and only if $P_d=P_{\vec{\theta}}$ almost everywhere.
The distribution of the generated samples can be estimated by their frequency of occurrences.
In numerical simulation, the exact distribution can be obtained from the wave function.
We draw the variation of the KL divergence w.r.t. the number of epoches in Fig. \ref{fig:KLD1}.
The number of layers influences the performance of the quantum GAN. For $L=1,2$, the capacity of the generator is not enough for generating the target distribution. A large standard deviation means in some runs the generator can produce the target distribution, but in other runs it can generate only part of the valid BAS patterns. For $L=3$, the trained generator can generate the target distribution in most of the 30 runs. For $L=4,5,6$, the KL divergence always converges to zero, which demonstrates that a deeper generative quantum circuit has better representation power.

\begin{figure}
\centering
\includegraphics[width=4in]{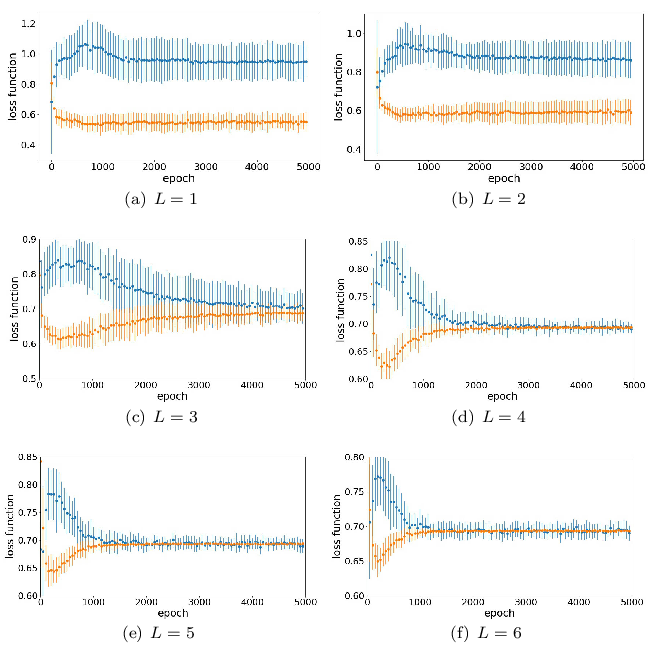}
\caption{Averages and standard deviations of the loss functions of the generator (in blue) and the discriminator (in orange) w.r.t. the number of epoches}
\label{fig:loss1}
\end{figure}

We also plot the loss functions of both the generator and the discriminator w.r.t. the number of epoches in Fig. \ref{fig:loss1}.
When the adversarial game reaches equilibrium, the output of the discriminator is 1/2 for both real and generated samples.
By substituting $D(x_i)=1/2$ into Eq. (\ref{eq:BCE}), we have $J_{final}=-\log\frac{1}{2}\approx 0.693$.
From Fig. \ref{fig:loss1} we can see that for $L=1,2$, the averages of two loss functions are separated. With the increase of $L$, they gradually converge to $J_{final}$.

\subsection{Numerical experiment 2}

In the second numerical experiment, the quantum generator is the MPS quantum circuit presented in section \ref{sec:generator2}.
We report 4 cases with $L$ and $V$ set to: (a) $L=2, V=1$, (b) $L=2, V=2$, (c) $L=2, V=3$, (d) $L=3, V=2$.
Because the amount of learnable parameters in the MPS circuit is $5NL(V+1)$, the number of parameters in these four cases is 80, 120, 160 and 180, respectively.
A model with more parameters can be regarded as having larger capacity.
For each case, we repeat the experiment 30 times and report the averages and standard deviations every 50 epoches.

\begin{figure}
\centering
\includegraphics[width=4in]{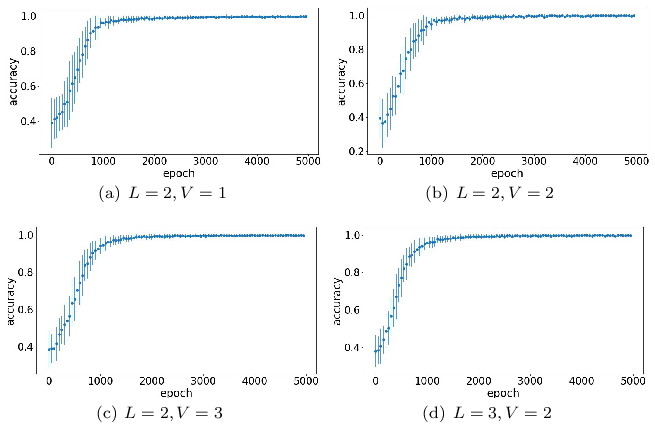}
\caption{Averages and standard deviations of the accuracy w.r.t. the number of epoches}
\label{fig:accuracy2}
\end{figure}

\begin{figure}
\centering
\includegraphics[width=4in]{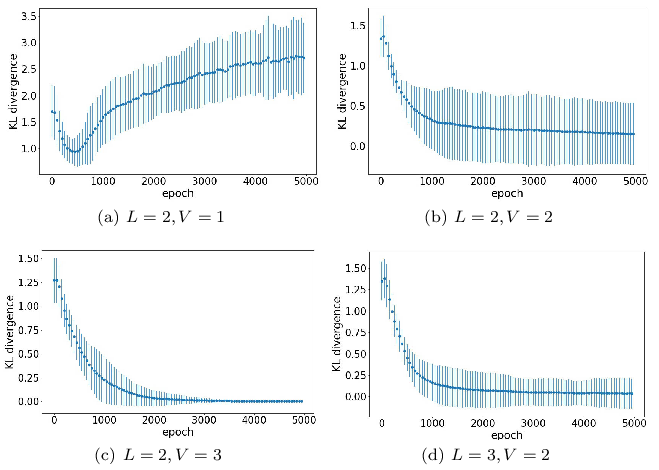}
\caption{Averages and standard deviations of the KL divergence w.r.t. the number of epoches}
\label{fig:KLD2}
\end{figure}

\begin{figure}
\centering
\includegraphics[width=4in]{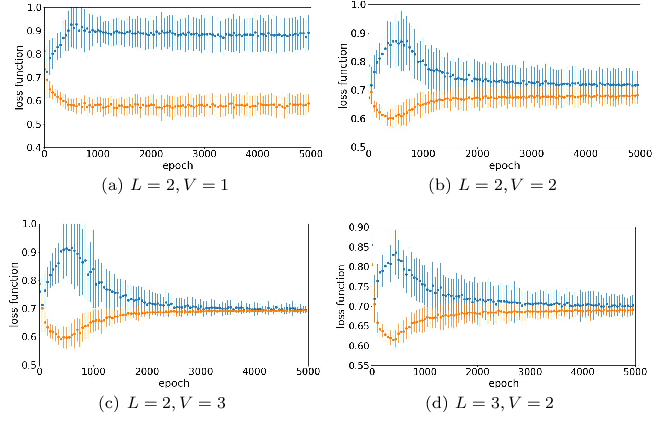}
\caption{Averages and standard deviations of the loss functions of the generator (in blue) and the discriminator (in orange) w.r.t. the number of epoches}
\label{fig:loss2}
\end{figure}

The generator accuracy w.r.t. the number of epoches is depicted in Fig. \ref{fig:accuracy2},
which shows that the accuracy increases very quickly and achieves nearly $100\%$ after 1000 epoches.
The variation of the KL divergence is depicted in Fig. \ref{fig:KLD2}.
We can see that the generated distribution gradually approaches the real data distribution with the increase of the capacity of the generator.
The variation of the loss functions of both the generator and the discriminator w.r.t. the number of epoches is plotted in Fig. \ref{fig:loss2}.
We can see that both loss functions converge to $J_{final}$ when the KL divergence approaches zero.

\section{Discussion and future work}\label{sec:discussion}
Our work is a combination of GANs and quantum computing, which can deal with the vanishing gradient problem, and has some other advantages. The proposed quantum GAN has a classical-quantum hybrid architecture which receives its input in the classical form by the classical discriminator and  outputs the discrete data by performing a measurement on the state generated from the quantum circuit. Thus it avoids the input/output bottlenecks embarrassing most of the existing quantum learning algorithms. Note that the classical-quantum hybrid architectures have recently attracted much attention, since it has been considered to be promising in  the  noisy intermediate-scale quantum (NISQ) era.
Furthermore, our quantum GAN can load the probability distribution implicitly given by data samples into a quantum state, which may be useful for some further applications.

Interesting future research directions include generating discrete data with higher dimension, choosing the layout of the generative quantum circuit, modeling the generator with non-unitary quantum dynamics, employing variants of GAN framework, using more heuristics to guide the training, and in-depth theoretical analysis of quantum GANs.
Our scheme adopts the approach of artificial neural network, which is a member of the soft computing family. We think it's interesting to combine quantum computing with other soft computing approaches such as genetic algorithms and fuzzy logic \cite{OZ14,AAA14,A17}.

\section{Conclusion}\label{sec:conclusion}
Researches on quantum versions of GANs seem very promising, due to the great potentiality and applications of GANs in machine learning and the advantage of quantum computing. The vanishing gradient problem implies that classical GANs are not good at producing discrete data.
In this paper, we propose a quantum GAN for generation of classical discrete data, which relies on a delicate way to estimate the analytic gradient of the quantum generator by sampling the same quantum generator with different parameters. By numerical simulation, we show our quantum GAN can generate simple BAS data distribution effectively.
A deeper quantum circuit has better capacity in generating discrete data. The accuracies of all the quantum GANs quickly converge to 1, which indicates that it's not difficult for the generator to learn to avoid producing non-BAS patterns. However, it is more difficult to generate patterns which are the same as that of the training dataset and thus more layers are needed.
The novel quantum GAN for discrete data generation can be regarded as a complement to classical GANS and deserve further research.

\end{document}